\documentclass[cits]{PoS}

\title{Properties of the Molecular Gas in Starburst Galaxies and AGN}

\ShortTitle{Properties of the Molecular Gas in Starburst Galaxies and AGN}

\author{\speaker{S.\ M\"uhle}\\
        Joint Institute for VLBI in Europe \\
        Postbus 2, 7990 AA Dwingeloo, The Netherlands\\
        E-mail: \email{muehle@jive.nl}}

\author{E.R.\ Seaquist\\
        Department of Astronomy and Astrophysics, University of Toronto\\
        50 St. George Street, Toronto, ON M5S 3H4, Canada\\
        E-mail: \email{seaquist@astro.utoronto.ca}}

\author{C.\ Henkel\\
        Max-Planck-Institut f\"ur Radioastronomie\\
        Auf dem H\"ugel 69, 53121 Bonn, Germany\\ 
        Email: \email{p220hen@mpifr-bonn.mpg.de}}

\abstract{
There is growing evidence that the properties of the molecular gas in the
nuclei of starburst galaxies and in AGN may be very different from those seen
in Galactic star forming regions and that a high kinetic temperature in the
molecular gas may lead to a non-standard initial mass function in the next
generation of stars. Unfortunately, among the fundamental parameters derived
from molecular line observations, the kinetic temperature of the molecular gas
in external galaxies is often not well determined due to a lack of suitable
tracer molecules. We discuss the diagnostic power of selected transition lines 
of formaldehyde (H$_2$CO), which can be used as a molecular thermometer as 
well as an excellent tracer of the molecular gas density. As a proof of 
concept, we present the results of our multi-transition line study of the 
H$_2$CO emission from the prototypical starburst galaxy M82. Using our large 
velocity gradient model, we tightly constrain the physical properties of the 
dense gas in the prominent molecular lobes, completely independent of the 
standard "cloud thermometer" ammonia (NH$_3$) or other molecular tracers. Our 
results agree well with the properties of the high-excitation molecular gas 
component found in the most comprehensive CO studies. Our observations also 
indicate that there may be an asymmetry between the two molecular lobes.
}

\FullConference{The 9th European VLBI Network Symposium on The role of VLBI in the Golden Age for Radio Astronomy and EVN Users Meeting\\
		 September 23-26, 2008\\
		 Bologna, Italy}

\newcommand{\arcsec}{$^{\prime\prime}$}
\newcommand{\htco}{H$_2$CO}
\newcommand{\kms}{km\,s$^{-1}$}
\newcommand{\farcs}{\mbox{$.\!\!^{\prime\prime}$}}

\begin{document}

\section{Introduction}

\begin{figure}
 \begin{center}
  \includegraphics[width=0.9\textwidth]{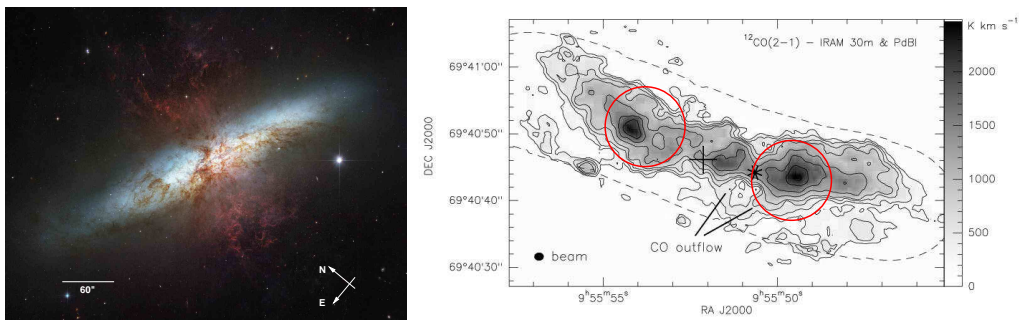}
 \end{center}
\caption{Left: Hubble Heritage image of the starburst galaxy M82 [14]. 
Right: High-resolution CO($2 \to 1$) map of the circumnuclear molecular ring 
   in M\,82 [20]. The observed pointing positions of our
   study and the corresponding beam widths at 218\,GHz (11\arcsec) are marked 
   by the solid circles.}
\label{fig1}
\end{figure}

Star formation, one of the fundamental processes that shape the evolution of a
galaxy and its surroundings, is thought to be fuelled by molecular gas, in 
particular by its dense component. Several observational studies have already 
shown the correlation between the dense molecular gas and star formation 
[e.g.\ 4]. 
But does star formation proceed in the same manner in all types of galaxies or 
does it depend significantly on the physical properties of the molecular gas? 
In recent years, more and more evidence has been found for the presence of a 
significant, or even dominant, warm molecular gas component in starburst 
galaxies and AGN [10,17] and for nonstandard initial mass functions in active 
environments [16,6]. Unfortunately, the physical properties of the molecular 
gas in external galaxies, in particular the kinetic temperature, are rarely 
well constrained. 

The easily thermalized and optically thick CO($1 \to 0$) and CO($2 \to 1$) 
transitions would be a good temperature tracer, if the filling factor of 
extragalactic clouds was better known. The inversion lines of the symmetric
top molecule ammonia (NH$_3$) are frequently used as ``cloud thermometer'' in 
our Galaxy.
However, variations in the fractional 
abundance of up to three orders of magnitude within the disk of the Milky Way 
may indicate that ammonia traces only a specific component of the molecular 
gas. Other commonly observed molecules 
like HCN are good density tracers, but require an a priori knowledge of the 
kinetic temperature. Therefore, the dust temperature is often {\it assumed} to 
be a good measure of the kinetic temperature of the gas.  

In this contribution, we present the results of our recent study on the
physical properties of the molecular gas in the prototypical nearby starburst 
galaxy M\,82, demonstrating the diagnostic power of selected para-formaldehyde 
lines for external galaxies [11]. Formaldehyde (H$_2$CO) is a 
slightly asymmetric top molecule with a wealth of millimeter and submillimeter 
transitions in its subspecies para-\htco\ and ortho-\htco\ and 
shows only little variation in its fractional abundance in a variety of 
galactic environments [e.g.\ 5]. In general, line 
intensity ratios involving different $K_a$ ladders of energy levels are 
good tracers of the kinetic temperature, because the relative 
populations of these ladders are governed by collisions. By contrast, line 
ratios within a single $K_a$ ladder sensitively probe the gas density, once 
the kinetic temperature is known [7]. Thus, H$_2$CO is
not only one of the few direct molecular thermometers, but also an excellent 
tracer of the gas density.

\section{The Properties of the Molecular Gas in M82}

\begin{figure}
 \begin{center}
  \includegraphics[width=0.9\textwidth]{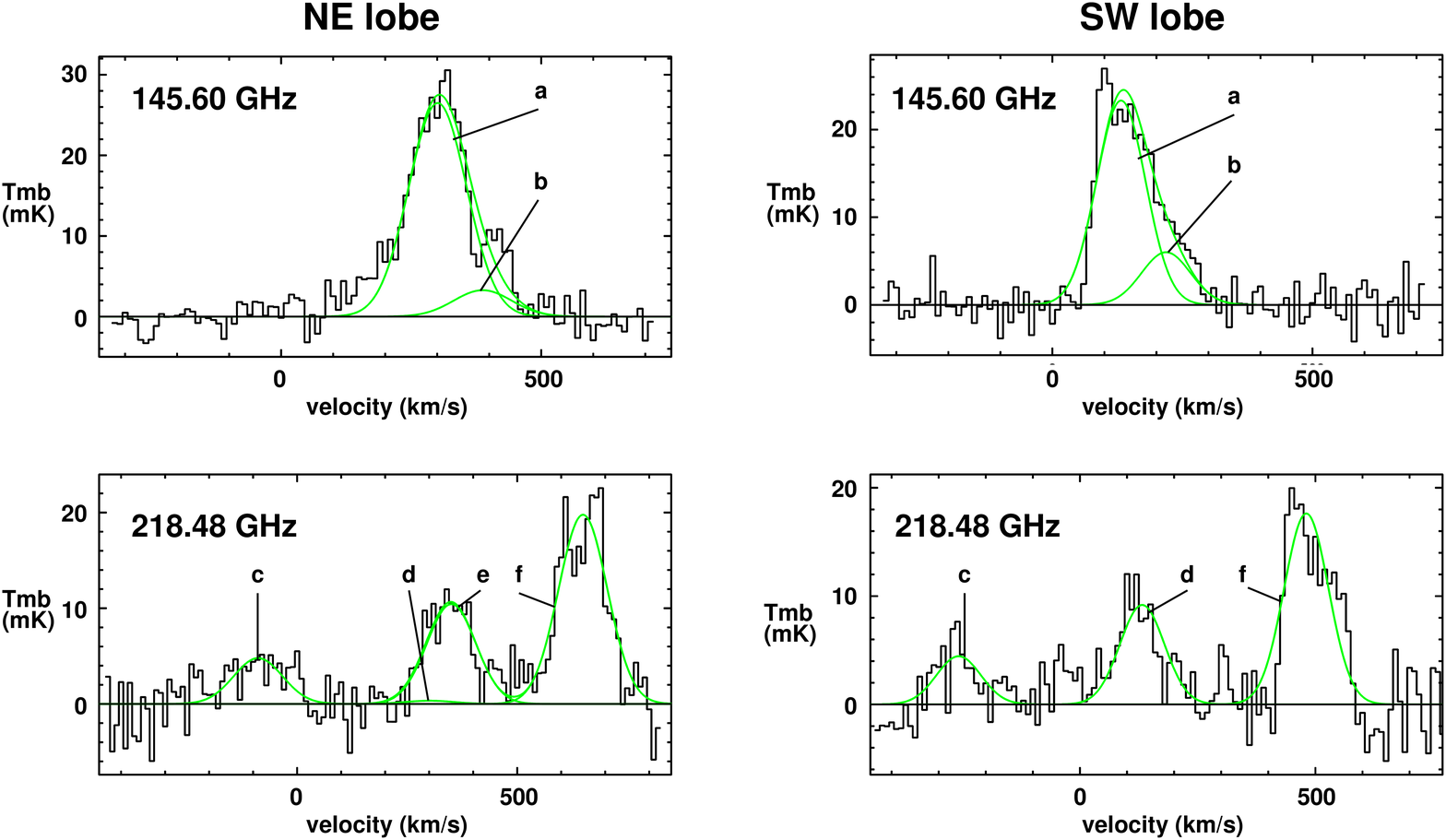}
 \end{center}
\caption{Observed spectra of the NE (left) and the SW lobe (right) of 
M\,82. Each spectrum is labeled with the frequency the receiver was tuned to.
Thus, the velocity scale of the 218.48\,GHz spectra refers to the 
\htco($3_{22} \to 2_{21}$) line. The \htco($3_{03} \to 2_{02}$) and the 
\htco($3_{21} \to 2_{20}$) transitions are offset by 348.5\,\kms\ and 
$-389.7$\,\kms, respectively, while the CH$_3$OH($4_2 \to 3_1\, E$) line is 
offset by 49.4\,\kms.  In the 146\,GHz spectra, the HC$_3$N($16 \to 15$) line 
is offset by  86.46\,\kms\ from the \htco($2_{02} \to 1_{01}$) emission. The 
curves show the Gaussian fit to each individual line as well as the spectrum 
resulting from a superposition of all identified lines: a = \htco($2_{02} \to
1_{01}$), b = HC$_3$N($16 \to 15$), c = \htco($3_{21} \to 2_{20}$), d =
\htco($3_{22} \to 2_{21}$), e = CH$_3$OH($4_2 \to 3_1\, E$), f =
\htco($3_{03}\to 2_{02}$).
}
\label{fig2}
\end{figure}

As the most prominent nearby galaxy in the northern hemisphere with a nuclear
starburst, M\,82 has been the target of many investigations, including
numerous molecular line studies [e.g.\ 3,18]. Figure 1 shows 
an optical image of the highly inclined
galaxy and its prominent galactic wind outlined in the H$\alpha$ emission
(red). The molecular gas is concentrated in a circumnuclear ring around the 
center of the starburst, which is seen nearly edge-on. The target of our 
observations were the two lobes of the ring, which are easily spatially 
separable with single-dish telescopes. Tuning the receivers
of the IRAM 30-m telescope to 145.60\,GHz and 218.48\,GHz, we detected the 
H$_2$CO transitions $2_{02} \to 1_{01}$, $3_{03} \to 2_{02}$, 
$3_{22} \to 2_{21}$, and $3_{21} \to 2_{20}$ at 145.60\,GHz, 218.22\,GHz, 
218.48\,GHz, and 218.76\,GHz, respectively (Fig.\ 2). Note that the latter 
three lines were observed simultaneously in a 1GHz wide spectrum, which 
eliminated uncertainties due to pointing errors, calibration issues or 
different beamwidths in the temperature-tracing line ratio 
\htco($3_{03} \to 2_{02}$)/\htco($3_{21} \to 2_{20}$). The HC$_3$N($16 \to
15$) line is also present in both 145.60\,GHz spectra, but easily 
separated from the nearby \htco\ line. Our observations also revealed
CH$_3$OH($4_2 \to 3_1\, E$) emission in the spectrum of the northeastern 
lobe. The presence of methanol in the molecular lobes of M\,82 was 
independently confirmed by [9]. 

The basic assumption that the observed emission lines all originate from the 
same volume of molecular gas with the same average physical properties implies 
that all lines should have the same velocity profile, parametrized here by a 
Gaussian curve with the same central velocity and the same line width for all
lines. 
With these constraints, we fitted all detected lines of a spectrum 
simultaneously using the technique of ``fitting dependent lines'' in 
{\tt class}, a part of the GILDAS software package. Accounting for the small 
difference in
beam widths between the observations at 145.60\,GHz and those at 218.48\,GHz, 
we derived the integrated line intensities of the detected \htco\ lines and 
compared the line intensity ratios of our observations with the results of our 
non-LTE models. These models are based on the LVG approximation and a
spherical cloud geometry and cover a wide parameter space in kinetic 
temperature, gas density and para-\htco\ column density per velocity interval 
(see [11] for 
details). The comparison suggests that the average kinetic temperature of the
molecular gas traced by the observed \htco\ lines is about 200\,K in both
lobes, a moderate gas density of $7 \cdot 10^3$\,cm$^{-3}$ and a total
molecular gas mass of about $1.5 \cdot 10^8\,M_{\odot}$ per lobe (Fig. 3). 
While the exact numbers depend slightly on the \htco\ abundance and the exact 
extent of the emission region, the kinetic temperature of the molecular gas is 
in any case much higher than the dust temperature of 48\,K [2].

\begin{figure}
\includegraphics[width=1.0\textwidth]{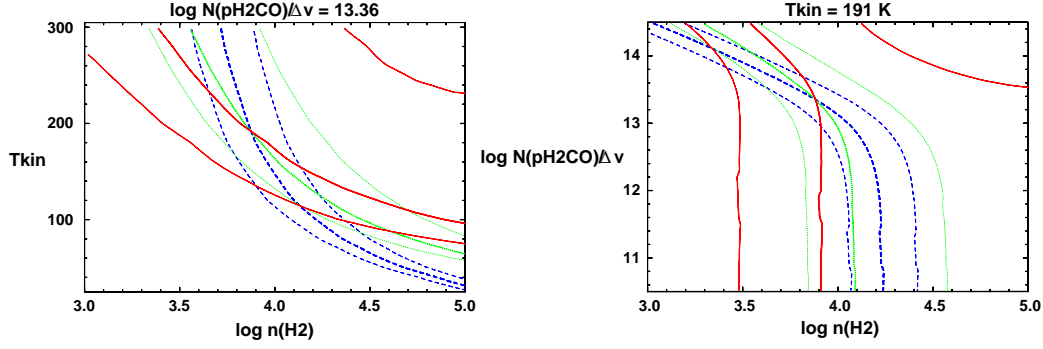}
\caption{Results of our LVG analysis for the NE and SW lobes shown as 
cuts through the 3-D parameter space ($T_{\rm kin}$ in K, $n_{\rm H2}$ in 
cm$^{-3}$, $N_{\rm pH2CO}/\Delta v$ in cm$^{-2}$\,km$^{-1}$\,s) along a plane 
of constant para-\htco\ column 
density per velocity interval (left) and of constant kinetic temperature 
(right). The lines represent the following line ratios: 
\htco($3_{03} \to 2_{02}$)/\htco($3_{21} \to 2_{20}$) (red), 
\htco($2_{02} \to 1_{01}$)/\htco($3_{03} \to 2_{02}$) (blue), 
\htco($2_{02} \to 1_{01}$)/\htco($3_{21} \to 2_{20}$) (green). The
corresponding dashed lines outline the uncertainties of each line ratio. The 
adopted source size is $\theta_s=7\farcs5$ and the assumed para-\htco\ 
abundance per velocity gradient is 
$\Lambda = 1 \times 10^{-9}\,{\rm km}^{-1}\,{\rm s\,pc}$.  }
\label{fig3}
\end{figure}

At first, a kinetic gas temperature of about 200\,K may seem odd considering 
that
the dust temperature is significantly lower. However, recent comprehensive 
studies of multiple CO lines suggest the presence of at least two distinct
molecular gas components in M\,82, where the highly excited gas component
may be the dominant one [19,8]. In these studies, the 
physical properties of the high-excitation molecular gas phase are remarkably
similar to our results, which were derived completely independently.   
Evidence for a warm molecular gas component has also been found in the
analysis of highly excited ammonia lines in nearby starburst galaxies 
[10] and in the investigation of IR rotational H$_2$
lines in a sample of starburst and Seyfert galaxies [17].
Shocks and strong UV and/or X-ray irradiation and cosmic-ray heating are among
the heating mechanisms that are likely to be found in active environments like
starbursts and AGN and that can heat a large fraction of the molecular gas to
temperatures of about 150\,K [1,17].

\section{Conclusions and Outlook}

\begin{figure}
\includegraphics[width=1.0\textwidth]{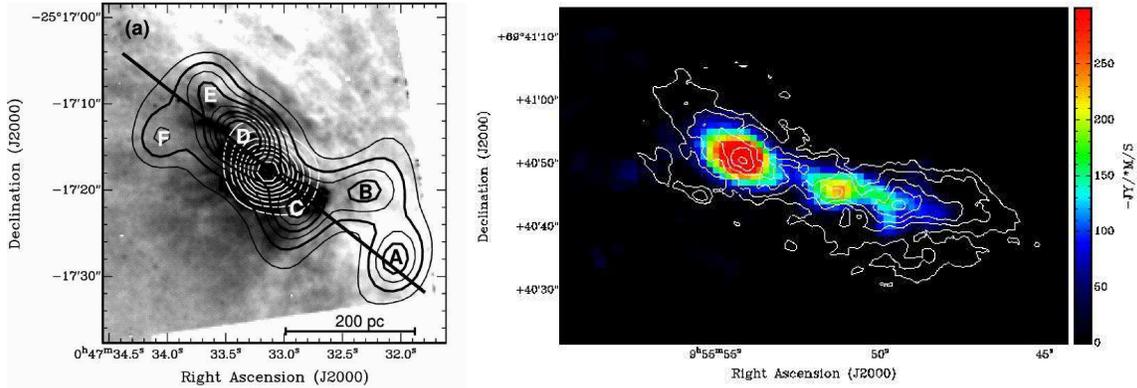}
\caption{Left: Superresolved NH$_3$(3, 3) emission displayed as black contours 
on a logarithmic HST WFPC2 F814W image of the core of NGC 253 [15]. 
Right: Preliminary map of the H$_2$CO($1_{10} \to 1_{11}$) line 
seen in absorption in the molecular ring of M82. The white contours outline the
CO($2 \to 1$) emission as mapped by [20].}
\label{fig4} 
\end{figure}

\begin{enumerate}
\item We have demostrated the diagnostic power of selected formaldehyde lines
    in deriving the physical properties of the molecular gas in external
    galaxies. 
\item We detected several para-\htco\ lines at 146\,GHz and at 218\,GHz in both
  molecular lobes of M\,82, including \htco\ lines of the $K_a$=2 ladder.
\item Our non-LTE line ratio analysis suggests the presence of warm 
  ($\sim 200$\,K), moderately dense ($7 \cdot 10^3$\,cm$^{-3}$) molecular gas 
  near the nuclear starburst, which is in very good agreement with the
  properties of the high-excitation molecular gas component found in recent 
  comprehensive studies of multiple CO lines.
\item Our case study supports the view that in active
  environments like starburst galaxies and AGN, a large fraction of the 
  molecular gas may have a kinetic temperature of a few hundred Kelvin. 
\item We find strong evidence for the presence of methanol in at least the
  northeastern lobe in M\,82, confirming the results of [9].
\end{enumerate}

Due to the instrumental limitations of the current telescopes and correlators,
there are still very few investigations of the properties of the
molecular gas in external galaxies at resolutions $<10$\arcsec\ using sensitive 
temperature probes. NGC\,253, a nearby starburst galaxy similar to M\,82, has 
been observed with the ATCA in ammonia inversion lines (Fig.\ 4), which suggest
kinetic temperatures of 140\,K in the northeastern molecular cloud complexes 
and of 200\,K in the southwestern complexes [15]. Preliminary results of 
high-resolution observations of M\,82 centered on the \htco($1_{10} \to 1_{11}$)
 line (Fig.\ 4) show the transition in {\em absorption} against the continuum 
[12]. The \htco\ morphology and kinematics seem to follow the distribution and 
kinematics of the CO emission [20] with the exception of the southwestern lobe,
which shows less \htco\ absorption than the northeastern lobe.

With the commissioning of ALMA and the new correlator of the
EVLA, we will be able to detect molecules like formaldehyde and ammonia with 
high spatial and frequency resolution in 
a large number of galaxies, employing their full diagnostic power to derive the
physical properties of the molecular gas in galactic environments very 
different from the Milky Way disk and thus to reveal the interdependence of 
molecular gas, star formation and feedback processes.

\acknowledgments
 We wish to thank the staff at Pico Veleta/Granada 
for their support during the observations. 
E.R.S. acknowledges a Discovery Grant from NSERC.  
This work has made use of the following software and resources: GILDAS, specx, 
Statistiklabor (FU Berlin, CeDiS), asyerr (Seaquist \& Yao), 
the JPL Catalog of spectral lines, The Cologne Database for Molecular 
Spectroscopy [12] and NASA's Astrophysics Data System 
Bibliographic Services (ADS).

\end{document}